\documentclass[aoas,seceqn,nameyear,rotating,dvips]{arximspdf}
\usepackage{dcolumn}

\doi{10.1214/10-AOAS377}
\volume{5}
\issue{1}
\pubyear{2011}
\firstpage{468}
\lastpage{485}

\makeatletter
\newproclaim{Algorithm}{Algorithm}
\newproclaim{example}{Example}
\newcolumntype{d}[1]{D{.}{.}{#1}}
\makeatother

\begin{document}
\begin{frontmatter}

\title{Random lasso}
\runtitle{Random lasso}

\begin{aug}
\author[A]{\fnms{Sijian} \snm{Wang}\ead[label=e1]{swang@biostat.wisc.edu}},
\author[B]{\fnms{Bin} \snm{Nan}\thanksref{t2}\ead[label=e2]{bnan@umich.edu}},
\author[C]{\fnms{Saharon} \snm{Rosset}\thanksref{t3}\ead[label=e3]{saharon@post.tau.ac.il}}
\and
\author[D]{\fnms{Ji} \snm{Zhu}\corref{}\thanksref{t4}\ead[label=e4]{jizhu@umich.edu} }

\thankstext{t2}{Supported in part by NSF Grant DMS-07-06700.}
\thankstext{t3}{Supported in part by EU Grant MIRG-CT-2007-208019 and ISF Grant 1227/09.}
\thankstext{t4}{Supported in part by NSF Grants DMS-07-05532 and DMS-07-48389.}

\runauthor{Wang et al.}

\affiliation{University of Wisconsin, University of Michigan, Tel
Aviv University and University of Michigan}

\address[A]{S. Wang\\
Department of Biostatistics\\
University of Wisconsin\\
Madison, Wisconsin, 53792\\USA\\
\printead{e1}}

\address[B]{B. Nan\\
Department of Biostatistics\\
University of Michigan\\
Ann Arbor, Michigan, 48109\\USA\\
\printead{e2}}

\address[C]{S. Rosset\\
School of Mathematical Sciences\\
Tel Aviv University\\
Tel Aviv\\
Israel\\
\printead{e3}}

\address[D]{J. Zhu\\
Department of Statistics\\
University of Michigan\\
Ann Arbor, Michigan, 48109\\USA\\
\printead{e4}}

\end{aug}

\received{\smonth{6} \syear{2008}}
\revised{\smonth{6} \syear{2010}}

%
\begin{abstract}
We propose a computationally intensive method, the random lasso
method, for variable selection in linear models. The method consists
of two major steps. In step 1, the lasso method is applied to many
bootstrap samples, each using a set of randomly selected covariates.
A measure of importance is yielded from this step for each
covariate. In step 2, a similar procedure to the first step is
implemented with the exception that for each bootstrap sample, a
subset of covariates is randomly selected with unequal selection
probabilities determined by the covariates' importance. Adaptive
lasso may be used in the second step with weights determined by the
importance measures. The final set of covariates and their
coefficients are determined by averaging bootstrap results obtained
from step 2. The proposed method alleviates some of the limitations
of lasso, elastic-net and related methods noted especially in the
context of microarray data analysis: it tends to remove
highly correlated variables altogether or select them all, and
maintains maximal flexibility in estimating their coefficients,
particularly with different signs; the number of selected variables
is no longer limited by the sample size; and the resulting
prediction accuracy is competitive or superior compared to the
alternatives. We illustrate the proposed method by extensive
simulation studies. The proposed method is also applied to a
Glioblastoma microarray data analysis.
\end{abstract}

%

\end{frontmatter}

%

\section{Introduction}
\label{sec:intro}

Suppose the training data set consists of $n$ observations
$(\mathbf{x}_1, y_1),\ldots,(\mathbf{x}_i, y_i),\ldots,(\mathbf
{x}_n,y_n)$, where
$\mathbf{x}_i=(x_{i1},\ldots,x_{ip})'$ is a $p$-dimensional vector of
predictors and $y_i$ is
the response variable. We consider the following linear model in this article:
%
%
\begin{equation}\label{eq:reg}
y_i = \beta_1x_{i1} + \cdots+\beta_px_{ip} + \varepsilon_i,
\end{equation}
where $\varepsilon_i$ is the error term with mean zero. We assume that
the response and the predictors are mean-corrected, so we can
exclude the intercept term from model~(\ref{eq:reg}).

Our motivating application comes from the area of microarray
data analysis [Horvath et al. (\citeyear{HorvathEtAl06})], which embodies some of
the properties of the model (\ref{eq:reg}) in many
modern applications:

\begin{longlist}[1.]
\item[1.] In a typical microarray study, the sample size $n$ is
usually on the order of~10s, while the number of genes $p$ is
on the order of 1000s or even 10,000s.
For example, in the glioblastoma microarray gene expression
study of Horvath et al. (\citeyear{HorvathEtAl06}), the sample sizes of the two
data sets are 55 and 65, respectively, while the number of
genes considered in their analysis is 3600.

\item[2.] Microarray data analysis typically combines predictive
performance and model interpretation as its goals:
one seeks models which explain the phenotype of interest well,
but also identify genes, pathways, etc. that might
be involved in generating this phenotype.
\end{longlist}

Shrinkage in general, and variable selection in particular,
feature prominently in such applications.
Significantly decreasing the number of variables used in the
model from the original 1000's to a more manageable
number by identifying the most useful and predictive ones usually
facilitates both improved accuracy and interpretation.



Variable selection has been studied extensively in the literature;
see \citet{Breiman95}, \citet{Tibs96}, \citet{FanLi01},
\citet{ZouHastie05} and \citet{Zou06}, among many others. In
particular, the lasso method proposed by \citet{Tibs96} has gained
much attention in recent years.

The lasso criterion penalizes the $L_1$-norm of the regression
coefficients:
%
%
\begin{equation}\label{eq:lasso}
\min_{\beta}\sum_{i=1}^n \Biggl(y_i - \sum_{j=1}^p
\beta_jx_{ij} \Biggr)^2 + \lambda\sum_{j=1}^p|\beta_j|,
\end{equation}
where $\lambda$ is a nonnegative tuning parameter.
Owing to the singularity of the derivative of $L_1$-norm penalty at
$\beta_j=0$,
lasso continuously shrinks the estimated coefficients toward zero, and some
estimated coefficients will be exactly zero when $\lambda$ is
sufficiently large.



Although lasso has shown success in many situations, it has two
limitations in practice [Zou and Hastie (\citeyear{ZouHastie05})]:

\begin{longlist}[1.]
\item[1.] When the model includes several highly correlated variables,
all of which are related to some extent to the response variable,
lasso tends to pick only one or a few of them and shrinks the rest to
0. This may not be a desirable feature. For example, in microarray
analysis, expression levels of genes that share one common
biological pathway are usually highly correlated, and these genes
may all contribute to the biological process, but lasso usually
selects only one gene from the group. An ideal method should be able
to select all relevant genes, highly correlated or not, while
eliminating trivial genes.

\item[2.] When $p>n$, lasso can identify at most only $n$ variables
before it saturates. This again may not be a desirable feature
for many practical problems, particularly microarray studies,
for it is unlikely that only such a small number of genes are
involved in the development of a complex disease. A method that is
able to identify more than $n$ variables should be more desirable
for such problems.
\end{longlist}





Several methods have been proposed recently to alleviate these two
possible limitations of lasso mentioned above, including the
elastic-net [\citet{ZouHastie05}], the adaptive lasso [\citet{Zou06}],
the relaxed lasso [\citet{Meinshausen07}] and VISA
[\citet{RadchenkoJames08}]. In particular, \citet{ZouHastie05}
proposed the elastic-net method, a penalized regression with the
mixture of the $L_1$-norm and the $L_2$-norm penalties of the
coefficients:
%
%
\begin{equation}\label{eq:enet}
\min_{\beta}\sum_{i=1}^n \Biggl(y_i-\sum_{j=1}^p \beta_jx_{ij}\Biggr)^2 +
\lambda_1\sum_{j=1}^p|\beta_j| +
\lambda_2\sum_{j=1}^p\beta_j^2,
\end{equation}
where $\lambda_1$ and $\lambda_2$ are two nonnegative tuning
parameters. Similar to lasso, the elastic-net method also
simultaneously does automatic variable selection and continuous
shrinkage. Due to the nature of the $L_2$-norm penalty, that is, the
ridge regression penalty, the number of selected variables is no longer
limited by the sample size. However, the ridge penalty forces the
estimated coefficients of highly correlated predictors to be close
to each other. This feature helps to select or remove highly
correlated variables altogether if their coefficients are truly
close to each other, but it loses the ability of estimating
coefficients of highly correlated variables with different
magnitudes, particularly with different signs, which is not rare in
practical problems. As a simple illustrative example, eggs are rich
in both protein and cholesterol that have quite different effects to
human health. When we consider the impact of egg consumption to
human health, we have two highly correlated variables with opposite
effects. In this scenario, forcing the estimated coefficients of
protein and cholesterol to be the same will cause big biases, and is
not expected to have adequate prediction performance.

Another modification of the lasso method is the adaptive lasso
proposed by \citet{Zou06}, which penalizes the weighted $L_1$-norm of
the regression coefficients:
%
%
\begin{equation}\label{eq:alasso}
\min_{\beta}\sum_{i=1}^n \Biggl(y_i - \sum_{j=1}^p \beta_jx_{ij}\Biggr)^2
+ \lambda\sum_{j=1}^p w_j |\beta_j|,
\end{equation}
where $w_j = |\hat\beta_j^{\mathrm{ols}}|^{-r}$ for a constant $r>0$,
and $\hat\beta_j^{\mathrm{ols}}$ is the classical ordinary least squares
(OLS) estimator for $\beta_j$. Adaptive lasso possesses some nice
asymptotic properties that lasso does not have. When $p$ is fixed,
$n$ tends to $\infty$ and $\lambda$ approaches zero with a
certain rate,
\citet{Zou06} has shown that the adaptive lasso approach selects the
true underlying model with probability tending to one, and the
corresponding estimated coefficients have the same asymptotic normal
distribution as they would have if the true underlying model were
provided in advance. This is called the ``oracle'' property by
\citet{FanLi01}, a property of super-efficiency. Although adaptive
lasso has nice asymptotic properties, its finite sample performance
does not always dominate lasso because it heavily depends on the
precision of the OLS estimation. In his Table 2, Zou [(\citeyear{Zou06}), page
1424]
presented that the prediction performance of adaptive lasso can be
worse than lasso when the OLS estimation is more variable. In
practice, adaptive lasso suffers (sometimes more severely than
lasso) from the multicollinearity caused by large correlations among
covariates because OLS estimates are very unstable in this
situation. In addition, due to the intrinsic constraint of the
$L_1$-norm penalty, the number of variables selected by adaptive
lasso cannot exceed $n$.




In this article we propose a novel extension of the lasso method,
which we call the random lasso method. The proposed method can
handle highly correlated variables in a more flexible manner than
elastic-net, especially when their effects have different magnitudes
and signs, and also can select more variables than the sample size.
Our experiments below demonstrate that the combination of
variable selection quality, estimation accuracy, and prediction
quality offered by the random lasso is consistently competitive
with, and often significantly superior, to those of all the
approaches mentioned above. The main price one pays for using the
random lasso, however, is in significantly increased computational
complexity.

The rest of the paper is organized as follows. We
introduce the proposed random lasso method in Section \ref{sec2}, and
demonstrate the method using
simulation studies in Section \ref{sec3}. In Section \ref{sec4} we analyze a real
data example, and in Section \ref{sec5} we provide a summary of the
proposed method.

\section{Random lasso}\label{sec2}

As mentioned above, one of the limitations of lasso is that it can
select only one or a few of a set of highly correlated important
variables. If several independent data sets were generated from the
same distribution, then we would expect lasso to select
nonidentical subsets of those highly correlated important variables
from different data sets, and our final collection may be most, or
perhaps even all, of those highly correlated important variables by
taking a union of selected variables from different data sets. Such
a process may yield more than~$n$ variables, overcoming the other
limitation of lasso.

In practice, however, we have only a single data set at hand, and
splitting the available data set into small pieces is not an
efficient way of using data. The bootstrap may yield desirable
perturbations similar to that of multiple data sets. Because each
bootstrap sample may include only a subset of the highly correlated
variables, the bootstrap has the ability to break down some of the
correlations. Hence, for each bootstrap sample, we can randomly
select $q$ candidate variables, with $q\leq p$.
This becomes the basic idea of the proposed random lasso
approach that has a similar flavor to the random forest method; see
\citet{Breiman01}. We also acknowledge that Park and Hastie (\citeyear{ParkHastie08})
proposed using bootstrap to provide a measure of how likely the
predictors were to be selected and examine what other predictors
could have been preferred. An obvious idea is to build on Park and
Hastie's idea to construct a complete predictive modeling tool which
may be termed ``Bagged Lasso.'' Our algorithm may be considered a
more evolved and ``adaptive'' version of this idea. In the
experiments below we discuss the effects of this added complexity on
performance.

Our proposed algorithm is a two-step approach and is described
below. In each step, bootstrap samples are drawn to yield the
desired perturbation similar to that of multiple data sets. To give
the method the most flexibility, we allow different numbers of
randomly selected variables to be included in the model, that is, $q_1$
candidate variables are randomly selected in each bootstrap sample
of the first step, and $q_2$ candidate variables are randomly
selected in each bootstrap sample of the second step, where $q_1$
and $q_2$ are treated as two tuning parameters that can be chosen as
large as $p$.

\begin{Algorithm*}[(``Generate'' and ``Select'')]
\textit{Step} 1. Generating importance measures for all coefficients:
\begin{longlist}
\item[1a.]Draw $B$ bootstrap samples with size $n$ by sampling
with replacement from the original training data set.

\item[1b.] For the $b_1$th bootstrap sample, $b_1 \in\{1, \ldots,
B\}$, randomly select $q_1$ candidate
variables, and apply lasso to obtain estimators
$\hat\beta_j^{(b_1)}$ for $\beta_j$, $j=1,\ldots,p$. Estimators
are zero for coefficients of those unselected variables, either
outside the subset of $q_1$ variables, or excluded by lasso.

\item[1c.] Compute the importance measure of $x_j$ by $I_j =
|B^{-1}\sum_{b_1=1}^B \hat\beta_j^{(b_1)}|$.
\end{longlist}

\textit{Step} 2. Selecting variables.
\begin{longlist}
\item[2a.] Draw another set of $B$ bootstrap samples with size
$n$ by sampling with replacement from the original training
data
set.

\item[2b.] For the $b_2$th bootstrap sample, $b_2 \in\{1, \ldots,
B\}$, randomly select $q_2$ candidate
variables with selection probability of $x_j$ proportional to
its importance $I_j$ obtained in step 1c, and apply lasso (or
adaptive lasso) to obtain estimators $\hat\beta_j^{(b_2)}$ for
$\beta_j$, $j=1,\ldots,p$.
Estimators are zero for coefficients of those unselected
variables, either outside the subset of $q_2$ variables, or
excluded by lasso.

\item[2c.] Compute the final estimator $\hat\beta_j$ of $\beta_j$
by $\hat\beta_j = B^{-1}\sum_{b_2=1}^B \hat\beta_j^{(b_2)}$.
\end{longlist}
\end{Algorithm*}

In step 1c, we would like to generate an importance score for
each predictor to assist variable selection and coefficient
estimation in the second step. The average coefficient for
each predictor over all bootstrap samples is our choice as an
importance score.
The intuition is that, for an unimportant
variable, the estimated coefficients in different bootstrap
samples are likely to be small or even have different signs, so
the corresponding average will typically be close to zero.
For an important variable, however, the estimated coefficients
in different bootstrap samples are likely to be consistently
large, and the corresponding average is also large. Therefore, we
choose the absolute value of the average as the importance
score for each predictor.

In step 2b, there are several choices of weights if adaptive
lasso is applied, for example,
\[
w_j = 1/|\hat\beta^{\mathrm{ols}}_j|^r,\qquad w_j = 1/|\hat\beta^{\mathrm{ridge}}_j|^r\quad \mbox{or}\quad  w_j = 1/|\hat\beta^{\mathrm{uni}}_j|^r,
\]
where $\hat\beta^{\mathrm{ols}}_j$ is the OLS estimator (if $p<n$),
$\hat\beta^{\mathrm{ridge}}_j$ is the ridge regression estimator,
$\hat\beta^{\mathrm{uni}}_j$ is the univariate estimator, and $r$ is a positive
number. Instead, we use importance measures obtained in step 1 as the
weights for adaptive lasso in our numerical examples and find it
works well.

In practice, we need to choose the number of bootstrap samples $B$,
the number of candidate variables to be included in each bootstrap
sample $q_1$ and $q_2$, and the tuning parameter $\lambda$ for
(adaptive) lasso to each bootstrap sample. Based on our experience,
our algorithm performs similarly when $B$ is large. One can take
$B=500$ or $B=1000$, for example. We can use cross-validation (CV)
to select~$q_1$ and $q_2$, and either CV or generalized
cross-validation (GCV) to select $\lambda$. In the following
simulations, we use independent validation data sets.


\section{Simulation studies}\label{sec3}

In this section we use simulations to demonstrate the proposed
random lasso method, and compare to a large collection of other
methods.
Data are generated from model (\ref{eq:reg}) with
$x_{ij}\sim N(0,1)$ and $\varepsilon_i \sim N(0,\sigma^2)$.

Five examples are considered. Examples \ref{ex1} and \ref{ex2} were used in the
lasso paper by \citet{Tibs96}, the adaptive lasso paper by
\citet{Zou06}, and the elastic-net paper by \citet{ZouHastie05}. In
Examples \ref{ex3} and \ref{ex4}, the coefficients of some highly correlated
variables have different signs. In Example \ref{ex5} the number of
variables with nonzero coefficients is larger than the sample size.
The following are the details of the five examples.

\begin{example}\label{ex1}
There are $p=8$ variables. The pairwise correlation
between $x_{j_1}$ and $x_{j_2}$ is set to be $\rho(j_1,j_2)=0.5^{|j_1-j_2|}$.
We let
\[
\beta= (3,1.5,0,0,2,0,0,0).
\]
Following Zou (\citeyear{Zou06}), we consider
three values of $\sigma$: $\sigma\in\{1,3,6\}$. The corresponding
signal-to-noise ratios (SNR) are $21.3$, $2.4$ and $0.6$,
respectively, where the SNR is defined as
$\operatorname{Var}(X'\beta)/\operatorname{Var}(\varepsilon)$.
\end{example}

\begin{example}\label{ex2}
We use the same model
in Example \ref{ex1} but with $\beta_j = 0.85$ for all~$j$. We also
consider the same three values of $\sigma$ as in Example \ref{ex1}. The
corresponding signal-to-noise ratios (SNR) are $14.5$, $1.6$ and $0.4$,
respectively.
\end{example}

\begin{example}\label{ex3}
There are $p=40$ variables. The first 10
coefficients are
nonzero. The correlation between each pair of the first 10
variables is set to be 0.9. The
remaining 30 variables are independent with each other, and also
independent with the first $10$ variables. We let
\[
\beta=(3,3,3,3,3,-2,-2,-2,-2,-2,0,\ldots,0),
\]
and $\sigma=3$. The
SNR is about $3.2$.
\end{example}

\begin{example}\label{ex4}
There are $p=40$ variables. The first six
coefficients are nonzero.
The pairwise correlation between the first three variables is set
to be $0.9$, and the same correlation structure is also set for the
second three
variables. The remaining $34$ variables are independent from each
other. The first three variables, the second three variables, and
the remaining $34$ variables are independent from each other. We let
\[
\beta=(3,3,-2,3,3,-2,0,\ldots,0),
\]
and $\sigma=6$. The SNR is
about $0.9$.
\end{example}

\begin{example}\label{ex5}
There are $p=120$ variables. The first 60 coefficients
are nonzero and drawn from $N(3, 0.5)$, and their values are then
fixed for all simulation runs. The remaining $60$ coefficients are set
to be zero. The covariate matrix is generated from a
multivariate normal distribution with zero mean and covariance
matrix as
\[
\pmatrix{
\Sigma_0 & 0 & 0 & 0 \vspace*{2pt}\cr
0 & \Sigma_0 & 0.2J & 0 \vspace*{2pt}\cr
0 & 0.2J & \Sigma_0 & 0 \vspace*{2pt}\cr
0 & 0 & 0 & \Sigma_0},
\]
where $\Sigma_0$ is a $30\times30$ matrix with unit diagonal
elements and off-diagonal elements of value $0.7$, and $J$ is a
$30\times30$ matrix with all unit elements.
\end{example}

For Examples \ref{ex1}--\ref{ex4}, we consider two sample sizes: $n=50$ and $n=100$.
For Example \ref{ex5}, since the purpose is to study the performance of
methods under the situation with $p>n$, we consider only sample size
$n=50$. For each example, we also generate a validation data set
with the same sample size as the training data set. Models are
fitted on training data only, and the validation data are used for
selecting the tuning parameters that minimize the prediction error
within their context respectively. Regarding the number of
bootstrap samples, we used $B=200$. We also tried $B=500$; the
results are similar to those of $B=200$.

We calculate the relative model error (RME) given below to evaluate
the prediction performance of each predictive model. Suppose that
the fitted coefficient vector is $\hat\beta$ and the true
coefficient vector is $\beta^0$, then the relative model error is
defined as follows:
\[
\mbox{Relative Model Error} = (\hat{\beta}-\beta^0)' \Sigma
(\hat{\beta}-\beta^0) / \sigma^2,
\]
where $\Sigma$ is the covariance matrix of the predictors, and
$\sigma$ is the standard deviation of the error term in model
(\ref{eq:reg}).

We repeat the simulation $100$ times and compute the average of RMEs
and their standard errors. We also record how frequently each
variable is selected during the $100$ simulations. For the variable
selection of random lasso, since the final estimator is the average
over all bootstrap samples, it is very easy for a variable to have a
nonzero coefficient if it has a nonzero coefficient in any
particular bootstrap sample. So it is a little unfair to use zero
or nonzero as the variable selection criterion for random lasso. In
this paper we introduce a threshold $t_n$, and consider a variable~$x_j$ to be selected, only if the corresponding coefficient
$|\hat\beta_j|>t_n$. In the following simulation studies, we chose
$t_n=1/n$, where $n$ is the sample size of the training data. We
compare the performance in prediction accuracy and variable
selection frequency of random lasso with the following methods: OLS,
lasso, adaptive lasso, elastic-net and two other recently developed
methods: relaxed lasso [(Meinshausen (\citeyear{Meinshausen07})] and VISA [Redchenko and
James (\citeyear{RadchenkoJames08})]. In Example \ref{ex5}, since $n<p$, the OLS estimator is not
unique, so we fitted ridge regression, and used the inverse of the
absolute value of the ridge regression estimator as the weight for
adaptive lasso. Results are summarized in Tables \ref{tab:sim1} and
\ref{tab:sim2}.

\begin{table}
\caption{$1000\times$ Average relative model errors of different
methods for all five examples}\label{tab:sim1}
\begin{tabular*}{\textwidth}{@{\extracolsep{\fill}}lcccccccc@{}}
\hline
&& \textbf{OLS} & \textbf{Lasso} & \textbf{ALasso} & \textbf{Enet} &
\textbf{Relaxo} & \textbf{VISA} & \textbf{RLasso}\\
\hline
\textit{Example} 1&&&&&&&& \\
$n=50$& $\sigma=1$ & 212 & 131 & 92 & 132 & \textit{88} & 91 & \textbf{82} \\
&& (13) & (9) & (8) & (9) & (7) & (8) & (7) \\
& $\sigma=3$ & 219 & 146 & 141 & 140 & \textit{101} & \textbf{97} & 105 \\
&& (11) & (9) & (9) & (9) & (7) & (7) & (7) \\
& $\sigma=6$ & 201 & 119 & 131 & \textit{104} & 117 & 115 & \textbf{96} \\
&& (11) & (6) & (7) & (6) & (7) & (7) & (6) \\
$n=100$& $\sigma=1$ & 89 & 59 & 37 & 59 & \textit{35} & 36 & \textbf{34} \\
&& (6) & (4) & (3) & (4) & (3) & (3) & (3) \\
&$\sigma=3$ & 90 & 59 & 48 & 58 & \textbf{38} & \textit{39} & 43 \\
&& (5) & (4) & (4) & (4) & (3) & (3) & (3) \\
& $\sigma=6$ & 89 & 56 & 59 & \textit{49} & 50 & 50 & \textbf{42} \\
&& (5) & (5) & (5) & (4) & (5) & (5) & (3) \\
\textit{Example} 2 &&&&&&&&\\
$n=50$& $\sigma=1$ & 218 & 211 & 229 & \textbf{181} & -- & \textit{210} & 217 \\
&& (13) & (12) & (13) & (10) & -- & (12) & (12) \\
& $\sigma=3$ & 202 & 171 & 200 & \textbf{140} & -- & 180 & \textit{167} \\
&& (10) & (8) & (10) & (8) & -- & (9) & (8) \\
& $\sigma=6$ & 203 & 127 & 158 & \textbf{111} & -- & 144 & \textit{112} \\
&& (12) & (5) & (7) & (6) & -- & (7) & (4) \\
$n=100$& $\sigma=1$ & 82 & 84 & 90 & \textbf{77} & -- & \textit{81} & 88 \\
&& (4) & (4) & (4) & (4) & -- & (4) & (4) \\
& $\sigma=3$ & 91 & 87 & 92 & \textbf{74} & -- & 95 & \textit{81} \\
&& (5) & (5) & (5) & (4) & -- & (6) & (7) \\
& $\sigma=6$ & 87 & 69 & 85 & \textbf{58} & -- & \textit{73} & 112 \\
&& (4) & (4) & (5) & (4) & -- & (4) & (4) \\
\textit{Example} 3 &&&&&&&&\\
 $n=50$ & &5259 & 666 & 613 & 562 & \textit{608} & 610 & \textbf{299} \\
& &(313) & (15) & (17) & (12) & (13) & (16) & (11) \\
 $n=100$& &680 & 505 & \textit{313} & 471 & 487 & 487 & \textbf{132} \\
& &(20) & (11) & (11) & (10) & (11) & (11) & (6) \\
\textit{Example} 4& &&&&&&&\\
$n=50$ & &4913 & 233 & 216 & 203 & 155 & \textit{152} & \textbf{126} \\
& &(323) & (11) & (12) & (10) & (9) & (9) & (5) \\
$n=100$&& 706 & 144 & 122 & 115 & 100 & \textit{98} & \textbf{70} \\
& &(25) & (6) & (5) & (5) & (5) & (5) & (36) \\
\textit{Example} 5& &&&&&&&\\
 $n=50$ & &\textbf{174} & 394 & 470 & 241 & 395 & 421 & \textit{227} \\
& &(9) & (12) & 11 & (11) & (11) & (12) & (11) \\
\hline
\end{tabular*}
\tabnotetext[]{}{\textit{Notes}: ALasso---the adaptive lasso
estimator; Enet---elastic net; RLasso---random lasso. The
numbers in the parentheses are the corresponding $1000\times$
standard errors. In each row, we mark the best performing method in
bold and the second best in italics.}
\end{table}
%

%
\begin{sidewaystable}
\tablewidth=\textwidth
\caption{Variable selection frequencies (\%) of different methods
for all five examples}\label{tab:sim2}
\begin{tabular*}{\textwidth}{@{\extracolsep{\fill}}lcccccc@{}}
\hline
& \textbf{Lasso} & \textbf{ALasso} & \textbf{Enet} & \textbf
{Relaxo} & \textbf{VISA}
& \textbf{RLasso}\\
\hline
\textit{Example} 1 &&&&&&\\
$n=50$ &&&&&&\\
\quad IV ($\sigma=1$) & (100, 100, 100) & (100, 100, 100) & (100, 100,
100) & (100, 100, 100) & (100, 100, 100) & (100, 100, 100) \\
\quad UV ($\sigma=1$) & (46, 58, 64) & (23, 27, 38) & (46, 59, 64) &
(10, 15, 19) & (11, 17, 20) & (28, 33, 44)\\
\quad IV ($\sigma=3$) & (99, 100, 100) & (95, 99, 100) & (100, 100,
100) & (93, 100, 100) & (97, 100, 100) & (99, 100, 100)\\
\quad UV ($\sigma=3$) & (48, 55, 61) & (33, 40, 48) & (44, 55, 69) &
(11, 18, 21) & (15, 21, 24) & (45, 57, 68) \\
\quad IV ($\sigma=6$) & (76, 85, 99) & (62, 76, 96) & (85, 92, 100) &
(60, 70, 98) & (61, 72, 98) & (92, 94, 100) \\
\quad UV ($\sigma=6$) & (47, 49, 53) & (32, 36, 38) & (43, 51, 70) &
(15, 19, 21) & (15, 19, 24) & (40, 48, 58) \\
$n=100$ &&&&&&\\
\quad IV ($\sigma=1$) & (100, 100, 100) & (100, 100, 100) & (100, 100,
100) & (100, 100, 100) & (100, 100, 100) & (100, 100, 100) \\
\quad UV ($\sigma=1$) & (54, 59, 64) & (27, 27, 32) & (53, 60, 63) &
(13, 16, 25) & (14, 20, 25) & (19, 29, 38)\\
\quad IV ($\sigma=3$) & (100, 100, 100) & (100, 100, 100) & (100, 100,
100) & (100, 100, 100) & (100, 100, 100) & (100, 100, 100)\\
\quad UV ($\sigma=3$) & (45, 51, 57) & (22, 30, 32) & (44, 55, 67) &
(6, 13, 17) & (13, 18, 19) & (36, 47, 56)\\
\quad IV ($\sigma=6$) & (96, 99, 100) & (86, 99, 100) & (99, 99, 100) &
(90, 93, 100) & (90, 93, 100) & (100, 100, 100) \\
\quad UV ($\sigma=6$) & (47, 57, 63) & (36, 40, 47) & (42, 63, 68) &
(11, 23, 25) & (11, 25, 27) & (37, 54, 61) \\[3pt]
\textit{Example} 2&&&&&& \\
$n=50$ &&&&&&\\
\quad IV ($\sigma=1$) & (100, 100, 100) & (100, 100, 100)& (100, 100,
100) & -- & (100, 100, 100) & (100, 100, 100)\\
\quad IV ($\sigma=3$) & (89, 92, 96) & (88, 90, 96) & (92, 96, 99) &
-- & (83, 88, 90) & (98, 99, 99)\\
\quad IV ($\sigma=6$) & (69, 72, 76) & (55, 60, 68) & (72, 78, 88) &
-- & (50, 54, 65) & (83, 89, 95)\\
$n=100$ &&&&&&\\
\quad IV ($\sigma=1$) & (100, 100, 100) & (100, 100, 100)& (100, 100,
100) & -- & (100, 100, 100) & (100, 100, 100)\\
\quad IV ($\sigma=3$) & (95, 97, 100) & (96, 97, 100) & (98, 99, 100) &
-- & (92, 95, 97) & (99, 100, 100) \\
\quad IV ($\sigma=6$) & (81, 86, 89) & (78, 81, 85) & (72, 78, 88) &
-- & (50, 55, 65) & (83, 89, 95) \\[3pt]
\hline
\end{tabular*}
\end{sidewaystable}
\setcounter{table}{1}
\begin{sidewaystable}
\tablewidth=\textwidth
\caption{(Continued)}
\begin{tabular*}{\textwidth}{@{\extracolsep{\fill}}lcccccc@{}}
\hline
& \textbf{Lasso} & \textbf{ALasso} & \textbf{Enet} & \textbf
{Relaxo} & \textbf{VISA}
& \textbf{RLasso}\\
\hline
\textit{Example} 3 &&&&&&\\
$n=50$ &&&&&&\\
\quad IV & (4, 35, 70) & (19, 38, 62) & (20, 60, 95) & (3, 29, 61) &
(2, 28, 60) & (93, 98, 100) \\
\quad UV & (14, 20, 30) & (6, 11, 18) & (6, 13, 18) & (7, 9, 15) &
(4, 7, 14) & (10, 17, 24) \\
$n=100$ &&&&&&\\
\quad IV & (45, 69, 95) & (68, 82, 93) & (51, 76, 99) & (39, 62, 88) &
(38, 62, 88) & (98, 99, 99) \\
\quad UV & (43, 52, 55) & (15, 21, 31) & (29, 35, 40) & (27, 36, 42) &
(27, 37, 43) & (22, 30, 37) \\[3pt]
\textit{Example} 4 &&&&&&\\
$n=50$ &&&&&&\\
\quad IV & (11, 70, 77) & (16, 49, 59) & (63, 92, 96) & (4, 63, 70) &
(4, 62, 73) & (84, 96, 97) \\
\quad UV & (12, 17, 25) & (4, 8, 14) & (9, 17, 23) & (0, 4, 9) & (1,
3, 8) & (11, 21, 30) \\
$n=100$ &&&&&&\\
\quad IV & (8, 84, 88) & (17, 62, 72) & (70, 98, 99) & (3, 75, 84) & (3, 76,
85) & (89, 99, 99) \\
\quad UV & (12, 22, 31) & (4, 10, 14) & (7, 14, 21) & (1, 3, 8) & (1,
4, 9) & (8, 14, 21) \\[3pt]
\textit{Example} 5 &&&&&&\\
\quad IV & (19, 30, 40) & (15, 25, 35) & (40, 50, 61) & (14, 23, 34) &
(16, 27, 35) & (76, 86, 95) \\
\quad UV & (3, 8, 14) & (0, 7, 11) & (1, 5, 8) & (0, 3, 8) & (0, 2,
8) & (18, 29, 38) \\
\hline
\end{tabular*}
\tabnotetext[]{}{\textit{Notes}: Since OLS always includes all variables, it
is excluded from the comparison. IV---important variables;
UV---unimportant variables. The three numbers in each pair of
parentheses are the min, median, and max of selection frequencies
among all important or unimportant variables, respectively.}
\end{sidewaystable}

As we can see from Table \ref{tab:sim1}, shrinkage methods perform
much better than OLS in most cases. This illustrates that some
regularization is crucial in achieving higher prediction accuracy.
We also see that random lasso has competitive RMEs with all other
methods in Examples \ref{ex1} and \ref{ex2}, except perhaps when compared to
elastic-net on Example \ref{ex2}. However, one should keep in mind that
Example \ref{ex2} represents the motivating setup for the elastic-net and,
thus, this result is not surprising. Random lasso has consistently
smaller RMEs than all other regularization methods in Examples \ref{ex3}--\ref{ex5}.
It also has the highest important variable selection frequency (see
Table \ref{tab:sim2}). In fact, random lasso selects most of the
important variables all the time. It also has competitive
performance in removing unimportant variables compared to other
methods in Examples \ref{ex1}, \ref{ex3} and \ref{ex4}. In Example \ref{ex5} random lasso selects
more unimportant variables than other methods, but it also selects
almost all important variables while other methods perform poorly on
this aspect.

It is interesting to compare the elastic net and the random lasso in
terms of the signs of the estimated nonzero coefficients of the
important variables in Examples~\ref{ex3} and~\ref{ex4}. In these two examples, the
important variables are highly correlated but with different signs.
The result is summarized in Tables \ref{tab:sign3} and
\ref{tab:sign4}. We can see that random lasso has much better
performance in estimating correct signs for truly negative
coefficients, and much smaller estimation bias than the elastic net
method.


%
\begin{table}[t]
\caption{Frequencies (\%) of the selected $q_1$ and $q_2$ for
Examples \protect\ref{ex1} and \protect\ref{ex2}} \label{tab:q1q2}
\begin{tabular*}{\textwidth}{@{\extracolsep{\fill}}lccccc@{}}
\hline
& $\bolds{q_2=2}$ & $\bolds{q_2=4}$ & $\bolds{q_2=6}$ & $\bolds{q_2=8}$ & \textbf{Total}\\
\hline
\multicolumn{6}{@{}l}{\textit{Example} 1: $n=50$}\\
$\sigma=1$&&&&&\\
\quad $q_1=2$ & \phantom{0}0 & \phantom{0}0 & \phantom{0}3 & \phantom{0}9 & \phantom{0}12\\
\quad $q_1=4$ & \phantom{0}0 & \phantom{0}2 & \phantom{0}3 & \phantom{0}8 & \phantom{0}13\\
\quad $q_1=6$ & \phantom{0}0 & \phantom{0}2 & \phantom{0}8 & \phantom{0}9 & \phantom{0}19\\
\quad $q_1=8$ & \phantom{0}0 & \phantom{0}7 & 21 & 28 & \phantom{0}56\\
\quad Total & \phantom{0}0 & 11 & 35 & 54 & 100\\[3pt]
$\sigma=3$&&&&&\\
\quad $q_1=2$ & \phantom{0}0 & \phantom{0}9 & 11 & 18 & \phantom{0}38\\
\quad $q_1=4$ & \phantom{0}0 & \phantom{0}2 & 10 & 10 & \phantom{0}22\\
\quad $q_1=6$ & \phantom{0}0 & \phantom{0}0 & \phantom{0}7 & \phantom{0}5 & \phantom{0}12\\
\quad $q_1=8$ & \phantom{0}0 & \phantom{0}1 & 12 & 15 & \phantom{0}28\\
\quad Total & \phantom{0}0 & 12 & 40 & 48 & 100 \\[3pt]
$\sigma=6$&&&&&\\
\quad $q_1=2$ & \phantom{0}8 & 22 & 15 & 17 & \phantom{0}62\\
\quad $q_1=4$ & \phantom{0}2 & \phantom{0}1 & \phantom{0}5 & \phantom{0}7 & \phantom{0}15\\
\quad $q_1=6$ & \phantom{0}0 & \phantom{0}1 & \phantom{0}6 & \phantom{0}6 & \phantom{0}13\\
\quad $q_1=8$ & \phantom{0}0 & \phantom{0}0 & \phantom{0}5 & \phantom{0}5 & \phantom{0}10\\
\quad Total & 10 & 24 & 31 & 35 & 100\\[6pt]
\multicolumn{6}{@{}l}{\textit{Example} 2: $n=50$}\\
$\sigma=1$&&&&&\\
\quad $q_1=2$ & \phantom{0}0 & \phantom{0}0 & \phantom{0}4 & 39 & \phantom{0}43\\
\quad $q_1=4$ & \phantom{0}0 & \phantom{0}0 & \phantom{0}1 & 30 & \phantom{0}31\\
\quad $q_1=6$ & \phantom{0}0 & \phantom{0}0 & \phantom{0}0 & 15 & \phantom{0}15\\
\quad $q_1=8$ & \phantom{0}0 & \phantom{0}0 & \phantom{0}1 & 10 & \phantom{0}11\\
\quad Total & \phantom{0}0 & \phantom{0}0 & \phantom{0}6 & 94 & 100\\[3pt]
$\sigma=3$&&&&&\\
\quad $q_1=2$ & \phantom{0}0 & 10 & 24 & 28 & \phantom{0}62\\
\quad $q_1=4$ & \phantom{0}0 & \phantom{0}1 & \phantom{0}2 & 11 & \phantom{0}14\\
\quad $q_1=6$ & \phantom{0}0 & \phantom{0}0 & \phantom{0}0 & 12 & \phantom{0}12\\
\quad $q_1=8$ & \phantom{0}0 & \phantom{0}0 & \phantom{0}1 & 11 & \phantom{0}12\\
\quad Total &\phantom{0}0 & 11 & 27 & 62 & 100\\[3pt]
$\sigma=6$&&&&&\\
\quad $q_1=2$ & \phantom{0}2 & 23 & 27 & 18 & \phantom{0}70\\
\quad $q_1=4$ & \phantom{0}0 & \phantom{0}1 & \phantom{0}2 & \phantom{0}5 & \phantom{00}8\\
\quad $q_1=6$ & \phantom{0}0 & \phantom{0}0 & \phantom{0}4 & \phantom{0}8 & \phantom{0}12\\
\quad $q_1=8$ & \phantom{0}0 & \phantom{0}0 & \phantom{0}4 & \phantom{0}6 & \phantom{0}10\\
\quad Total & \phantom{0}2 & 24 & 37 & 37 & 100\\
\hline
\end{tabular*}\vspace*{-3pt}
\end{table}
%

%
\begin{sidewaystable}
\tablewidth=\textwidth
\caption{Coefficient and coefficient sign estimation of elastic net
and random lasso for Example \protect\ref{ex3}} \label{tab:sign3}
\begin{tabular*}{\textwidth}{@{\extracolsep{\fill}}lcccccccccc@{}}
\hline
& \multicolumn{1}{c}{$\bolds{\beta_1}$} & \multicolumn{1}{c}{$\bolds{\beta_2}$} & \multicolumn{1}{c}{$\bolds{\beta_3}$}
& \multicolumn{1}{c}{$\bolds{\beta_4}$} & \multicolumn{1}{c}{$\bolds{\beta_5}$} &
\multicolumn{1}{c}{$\bolds{\beta_6}$} & \multicolumn{1}{c}{$\bolds{\beta_7}$} & \multicolumn{1}{c}{$\bolds{\beta_8}$}
& \multicolumn{1}{c}{$\bolds{\beta_9}$} & \multicolumn{1}{c@{}}{$\bolds{\beta_{10}}$}\\
\hline
True coef. & 3 & 3 & 3 & 3 & 3 & $-$2 & $-$2 & $-$2 & $-$2 & $-$2 \\[3pt]
Enet ($n=50$)&&&&&&&&&&\\
\quad Ave. of est. & 1.03 & 1.06 & 0.91 & 1.04 & 0.98 & $-$0.05 & $-$0.03 &
$-$0.03 & 0.01 & 0.04 \\
& (0.07) & (0.07) & (0.06) & (0.08) & (0.07) & (0.06) & (0.04) &
(0.05) & (0.03) & (0.02)\\
\quad Freq. (\%) of pos. sgn. & 94 & 91 & 92 & 95 & 91 & 23 & 16 & 17 & 19 &
27 \\
\quad Freq. (\%) of neg. sgn. & 0 & 0 & 0 & 0 & 0 & 5 & 6 & 3 & 4 & 1
\\[3pt]
RLasso ($n=50$)&&&&&&&&&& \\
\quad Ave. of est. & 1.84 & 2.01 & 1.75 & 1.81 & 1.84 & $-$0.84 & $-$0.89 & $-$0.88
& $-$0.91 & $-$0.83\\
& (0.12) & (0.12) & (0.12) & (0.11) & (0.11) & (0.09) & (0.07) & (0.07) &
(0.07) & (0.07)\\
\quad Freq. (\%) of pos. sgn. & 98 & 99 & 96 & 98 & 100 & 9 & 4 & 4 & 7 & 2\\
\quad Freq. (\%) of neg. sgn. & 2 & 0 & 2 & 2 & 0 & 88 & 95 & 93 & 93 &
97\\[3pt]
Enet ($n=100$)&&&&&&&&&&\\
\quad Ave. of est. & 1.42 & 1.54 & 1.47 & 1.43 & 1.61 & $-$0.53 & $-$0.52 &
$-$0.47 & $-$0.38 & $-$0.52\\
& (0.10) & (0.09) & (0.10) & (0.09) & (0.11) & (0.09) & (0.09)
& (0.09) & (0.07) & (0.09)\\
\quad Freq. (\%) of pos. sgn. & 98 & 99 & 98 & 97 & 98 & 15 & 20 & 17 & 17 &
19\\
\quad Freq. (\%) of neg. sgn. & 0 & 0 & 0 & 0 & 0 & 37 & 34 & 33 & 34 &
35\\[3pt]
RLasso ($n=100$)&&&&&&&&&& \\
\quad Ave. of est. & 2.33 & 2.51 & 2.45 & 2.31 & 2.48 & $-$1.51 & $-$1.35 &
$-$1.46 & $-$1.33 & $-$1.41\\
& (0.09) & (0.09) & (0.09) & (0.08) & (0.09) & (0.07) & (0.06)
& (0.07) & (0.06) & (0.07) \\
\quad Freq. (\%) of pos. sgn. & 99 & 99 & 99 & 98 & 99 & 1 & 0 & 1 & 1 & 0\\
\quad Freq. (\%) of neg. sgn. & 0 & 0 & 0 & 0 & 0 & 98 & 99 & 98 & 98 &
99\\
\hline
\end{tabular*}
\end{sidewaystable}
%
%
\begin{table}
\caption{Coefficient and coefficient sign estimation of elastic net
and random lasso for Example \protect\ref{ex4}} \label{tab:sign4}
\begin{tabular*}{\textwidth}{@{\extracolsep{\fill}}lcccccc@{}}
\hline
& \multicolumn{1}{c}{$\bolds{\beta_1}$} & \multicolumn{1}{c}{$\bolds{\beta_2}$} & \multicolumn{1}{c}{$\bolds{\beta_3}$} &
\multicolumn{1}{c}{$\bolds{\beta_4}$} & \multicolumn{1}{c}{$\bolds{\beta_5}$} &
\multicolumn{1}{c@{}}{$\bolds{\beta_6}$}\\
\hline
True coef. & 3 & 3 & $-$2 & 3 & 3 & $-$2 \\[3pt]
Enet ($n=50$)&&&&&&\\
\quad Ave. of est. & 1.30 & 1.44 & 0.51 & 1.75 & 1.47 & 0.74\\
& (0.07) & (0.08) & (0.06) & (0.09) & (0.07) & (0.07) \\
\quad No. of pos. sgn. & 92 & 94 & 63 & 96 & 92 & 70\\
\quad No. of neg. sgn. & 0 & 0 & 0 & 0 & 0 & 1\\[3pt]
RLasso ($n=50$) &&&&&&\\
\quad Ave. of est. & 1.85 & 1.68 & $-$0.17 & 2.01 & 1.89 & $-$0.17 \\
& (0.12) & (0.13) & (0.07) & (0.14) & (0.13) & (0.09) \\
\quad No. of pos. sgn. & 98 & 90 & 33 & 91 & 96 & 38\\
\quad No. of neg. sgn. & 1 & 7 & 65& 5 & 2 & 57\\[3pt]
Enet ($n=100$)&&&&&&\\
\quad Ave. of est. & 1.57 & 1.57 & 0.54 & 1.69 & 1.67 & 0.61 \\
& (0.06) & (0.07) & (0.05) & (0.06) & (0.06) & (0.05)\\
\quad No. of pos. sgn. & 97 & 98 & 69 & 98 & 99 & 72 \\
\quad No. of neg. sgn. & 0 & 0 & 0 & 0 & 0 & 0 \\[3pt]
RLasso ($n=100$)&&&&&& \\
\quad Ave. of est. & 2.25 & 1.91 & $-$0.57 & 2.28 & 2.08 & $-$0.55\\
& (0.06) & (0.07) & (0.05) & (0.06) & (0.06) & (0.05)\\
\quad No. of pos. sgn. & 99 & 97 & 17 & 100 & 99 & 15\\
\quad No. of neg. sgn. & 0 & 2 & 81 & 0 & 0 & 83\\
\hline
\end{tabular*}
\end{table}

For random lasso, the $q_1$ and $q_2$ selection can be
crucial. For Examples \ref{ex1} and \ref{ex2}, we select the optimal $q_1$ and $q_2$
based on the validation data set among values 2, 4, 6 and 8, for Examples
\ref{ex3} and \ref{ex4}, we select the optimal $q_1$ and $q_2$ among values 4, 8, 12,
16, 20, 24 and 28, and for Example \ref{ex5}, we select the optimal $q_1$ and
$q_2$ among values 5, 10, 15, 20 and 25.
We also summarize the frequency for the
selected $q_1$ and $q_2$ in Examples \ref{ex1} and \ref{ex2} with sample size
$n=50$ (see Table \ref{tab:q1q2}).
From the last columns and the last rows of the six sub-tables, we
can see that random lasso prefers a smaller number of predictors
in both the first stage and the second stage of the algorithm, as
$\sigma$ becomes larger (correspondingly, the
signal-to-noise ratio is smaller).
This illustrates that the random subset selection of
variables in each bootstrap sample can be helpful, when the
signal-to-noise ratio is small.

It should be noted that we also experimented with ``Bagged
Lasso'' (that is, a 1-step bootstrap approach with $q=p$) on all
simulations. The results were reasonable and, in fact, very similar
to the elastic net results on all setups. However, since these
results are clearly inferior overall to the random lasso, we chose
to eliminate them to avoid clutter.

%

\section{Glioblastoma gene expression data analysis}\label{sec4}
We analyze the data from a glioblastoma microarray gene expression
study conducted by Horvath et al. (\citeyear{HorvathEtAl06}) by using the proposed
random lasso method and compare with the lasso, adaptive lasso,
relaxed lasso, elastic-net and VISA methods.

%
\begin{table}[b]
\caption{Analysis of the glioblastoma data set}
\label{tab:realdata}
\begin{tabular*}{250pt}{@{\extracolsep{\fill}}lcc@{}}
\hline
\textbf{Method} & \textbf{No. of genes selected} & \textbf{Mean prediction error} \\
\hline
Lasso & 29 & 1.118 (0.205) \\
Adaptive lasso & 33 & 1.143 (0.211) \\
Relaxed lasso & 23 & 1.054 (0.194) \\
Elastic-net & 28 & 1.113 (0.204) \\
VISA & 15 & 0.997 (0.188) \\
Random lasso & 58 & 0.950 (0.210) \\
\hline
\end{tabular*}
\end{table}

Glioblastoma is the most common primary malignant brain tumor of
adults and one of the most lethal of all cancers. Patients with this
disease have a median survival of 15 months from the time of
diagnosis despite surgery, radiation, and chemotherapy. Global gene
expression data from two independent sets of clinical tumor samples
of $n=55$ and $n=65$ are obtained by high-density Affymetrix
arrays. Expression values of 3600 genes are available. Among the
first set of 55 patients, five were alive at the last followup and
four were alive in the second set. In our analysis, we exclude these
nine~censored subjects, and use the logarithm of time to death as
the response variable. The first data set is used as the training
set and the second data set as the test set.
\vadjust{\eject}

We first assess each of the 3600 genes by running simple linear
regression on the training set, and then select 1000 genes with the
smallest $p$-values. Starting with these 1000 genes, we fit a
linear regression model by the proposed random lasso method on the
training set, and select 58 genes. Table \ref{tab:genelist} lists
the gene symbol and estimated coefficient for each of these 58
genes. The model with these selected 58 genes is then used to
predict the log-survival times for subjects in the test set. We also
analyze the data using other lasso-related methods
starting with the same
1000 genes on the training set and evaluate obtained models using
the test set.

%
\begin{table}
\caption{Gene symbol and estimated coefficient for each of the 58
genes selected by random lasso based on 50 subjects in the
training set} \label{tab:genelist}
\begin{tabular*}{\textwidth}{@{\extracolsep{\fill}}ld{2.3}ld{2.3}@{}}
\hline
\textbf{Gene symbol} & \multicolumn{1}{c}{\textbf{Estimated coefficients}} &\multicolumn{1}{c}{\textbf{Gene symbol}} &
\multicolumn{1}{c@{}}{\textbf{Estimated
coefficients}} \\
\hline
VSNL1 & -3.839 &KIAA0194 & -0.039 \\
SNAP25 & -1.561 &MOBP & -0.033 \\
UBE2D3 & -0.382 &PTGDS & -0.028 \\
ARF4 & -0.341 &KIF5A & -0.024 \\
CSNK1A1 & -0.319 &GORASP2 & -0.021 \\
C13orf11 & -0.312 &ME2 & -0.019 \\
CHGA & -0.310 &CGI-141 & -0.019 \\
C11orf24 & -0.223 &p25 & -0.017 \\
OPTN & -0.221 &UGT8 & -0.016 \\
UNC84B & -0.176 &CKMT1 & -0.014 \\
S100A1 & -0.157 &KIF1A & -0.013 \\
KCNS1 & -0.155 &KCNAB2 & -0.012 \\
NPY & -0.124 &C3orf4 & -0.011 \\
TIP-1 & -0.107 &DNASE1L1 & -0.011 \\
FAIM2 & -0.086 &RNF44 & 0.011 \\
FSTL3 & -0.074 &ATP6V1B2 & 0.012 \\
NEFH & -0.072 &POLR3E & 0.012 \\
CTSK & -0.071 &LIN7C & 0.014 \\
RGS3 & -0.071 &GBP2 & 0.015 \\
PGCP & -0.070 &CSF1R & 0.018 \\
FLJ20254 & -0.059 &JIK & 0.019 \\
ANXA2 & -0.053 &--- & 0.019 \\
FLJ11155 & -0.052 &C1S & 0.026 \\
P2RX4 & -0.049 &ARHGAP15 & 0.040 \\
GPNMB & -0.044 &PPM1H & 0.063 \\
ICAM5 & -0.043 &MARK4 & 0.071 \\
ADIPOR1 & -0.043 &HPCAL4 & 0.196 \\
BSCL2 & -0.042 &SULT4A1 & 0.785 \\
AMBP & -0.042 &BSN & 2.662 \\
\hline
\end{tabular*}
\end{table}

Table \ref{tab:realdata} shows the number of genes selected by each
of these six methods in the training set and corresponding mean
prediction error in the test set. We can see that random lasso has the
smallest prediction error. It also selects more genes than the
other five methods. Among the 58 genes selected by random lasso, 7
genes are also selected by lasso, adaptive lasso, relaxed lasso, VISA and
elastic-net (for adaptive lasso, the adaptive weights were
calculated using ridge regression).

Several genes identified by the proposed method are of interest.
VSNL1, RGS3 and S100A4 are identified to be negatively associated
with the patients' survival. VSNL1 is a member of the
visinin/recoverin subfamily of neuronal calcium sensor proteins. The
encoded protein is strongly expressed in granule cells of the
cerebellum where it associates with membranes in a calcium-dependent
manner and modulates intracellular signaling pathways of the central
nervous system by directly or indirectly regulating the activity of
adenylyl cyclase. A previous study [Xie et al. (\citeyear{XieEtAl07})] has
demonstrated that VSNL1 plays a very important role in neuroblastoma
metastasis, and VSNL1 mRNA in highly invasive cells is significantly
higher than that in lowly invasive cells. RGS3 encodes a member of
the regulator of the G-protein signaling (RGS) family. This protein is a
GTP-ase activating protein which inhibits G-protein mediated signal
transduction. Tatenhorst et al. (\citeyear{Tatenhorst04}) demonstrated that glioma
cell clones overexpressing RGS3 showed an increase of both adhesion
and migration. S100A4 encodes a member of the S100 family of
proteins, which are localized in the cytoplasm and/or nucleus of a
wide range of cells, and involved in the regulation of a number of
cellular processes such as cell cycle progression and
differentiation. It is conjectured that the protein encoded by
S100A4 may function in motility, invasion, and metastasis [Zou et
al. (\citeyear{ZouEtAl05})]. VSNL1, RGS3 and S100A4 were also identified to be
related to the poor survival of brain tumor patients in Freije et
al. (\citeyear{Freije04}). BSN is identified to be positively associated with the
patients' survival. This gene is expressed primarily in neurons in
the brain, and the protein encoded by this gene is thought to be a
scaffolding protein involved in organizing the presynaptic
cytoskeleton. Additional studies will be required to establish the
direct relationships between the expression of these genes and the
Glioblastoma tumor behavior.

It is also interesting to observe that estimated coefficients of
VSNL1 and BSN ($-$3.839 and 2.662, resp.) have different signs,
but the correlation between the expression levels for
these two genes in the training set is very high ($\rho=0.96$).
Neither lasso nor elastic-net picked up these two genes. It is worth
conducting more detailed experiments to further explore the
connection between VSNL1 and BSN, and their relations to the
Glioblastoma tumor behavior.

\section{Conclusion}\label{sec5}
We have proposed the random lasso method for variable selection. The
idea of random lasso is mimicking the random forest method [Breiman
(\citeyear{Breiman01})] for linear regression models. By drawing bootstrap samples
from the original training set and randomly selecting candidate
variables, the average of the predictive models based on multiple
bootstrap samples alleviates two possible limitations of lasso. It
tends to select or remove highly correlated variables more
efficiently and has more flexibility in estimating their
coefficients than the elastic-net method. The number of variables
selected by random lasso is not limited by the sample size.
Simulation studies show that the proposed random lasso method has
good prediction performance compared to a large set of competitor
approaches, and the analysis of Glioblastoma microarray data set
demonstrates the usefulness of the proposed method in practice.

\section*{Acknowledgments}
We would like to thank the Editor, an Associate Editor and
three reviewers for their thoughtful and useful comments.
%


\printaddresses

\end{document}